\documentclass[review]{elsarticle}

\usepackage{lineno,hyperref,amsmath,amssymb,multirow}
\usepackage{subcaption}
\journal{Journal of Neurocomputing Templates}

\begin{document}

\begin{frontmatter}

\title{Blur-Attention: A boosting mechanism for non-uniform blurred image restoration}
\tnotetext[mytitlenote]{The work in this paper is supported by the National Natural Science Foundation of China (No. 61471013 and No.61701011), the  Beijing Education Committee Cooperation Beijing Natural Science Foundation (KZ201810005002, KZ201910005007).}


\author[mymainaddress,mysecondaryaddress]{Xiaoguang Li\corref{mycorrespondingauthor}}
\cortext[mycorrespondingauthor]{Corresponding author}
\ead{lxg@bjut.edu.cn}


\author[mymainaddress]{Feifan Yang}

\author[mythirdaddress]{Kin Man Lam}

\author[mymainaddress,mysecondaryaddress]{Li Zhuo}

\author[mymainaddress,mysecondaryaddress]{Jiafeng Li}

\address[mymainaddress]{Faculty of Information Technology, Beijing University of Technology, Beijing 100124, China}
\address[mysecondaryaddress]{Beijing Key Laboratory of Computational Intelligence and Intelligent System, Beijing University of Technology, Beijing
100124, China}
\address[mythirdaddress]{Department of Electronic and Information Engineering, The Hong Kong Polytechnic University, Hong Kong.}
\begin{abstract}

Dynamic scene deblurring is a challenging problem in computer vision. It is difficult to accurately estimate the spatially varying blur kernel by traditional methods. Data-driven-based methods usually employ kernel-free end-to-end mapping schemes, which are apt to overlook the kernel estimation. To address this issue, we propose a blur-attention module to dynamically capture the spatially varying features of non-uniform blurred images. The module consists of a DenseBlock unit and a spatial attention unit with multi-pooling feature fusion, which can effectively extract complex spatially varying blur features. We design a multi-level residual connection structure to connect multiple blur-attention modules to form a blur-attention network. By introducing the blur-attention network into a conditional generation adversarial framework, we propose an end-to-end blind motion deblurring method, namely Blur-Attention-GAN (BAG), for a single image. Our method can adaptively select the weights of the extracted features according to the spatially varying blur features, and dynamically restore the images. Experimental results show that the deblurring capability of our method achieved outstanding objective performance in terms of PSNR, SSIM, and subjective visual quality. Furthermore, by visualizing the features extracted by the blur-attention module, comprehensive discussions are provided on its effectiveness.
\end{abstract}

\end{frontmatter}


\section{Introduction}
Motion blur is a common degradation in images or videos captured by surveillance cameras. In the process of image acquisition, due to the influence of camera shake, depth changes, and target movement, non-uniform blur often occurs in images. Blind motion deblurring is an ill-posed inverse problem with non-uniform blur kernels, which is a fundamental challenging problem in computer vision.

Traditional deblurring methods can be divided into two categories: non-blind restoration and blind restoration. The non-blind restoration methods assume that the blur kernel is known, and the image deblurring problem is addressed by a deconvolution problem. Handcrafted priors on natural images usually work as regularization constraints. The blind restoration methods assume that the blur kernel is unknown. These methods attempt to recover the clean latent image by accurately estimating the blur kernel of each pixel. This makes the ill-posed problem worse. It is difficult for traditional methods to accurately estimate the blur features for real motion-blurred images.

Due to the nature of being ill-posed for single image deblurring, traditional blind restoration methods \cite{bag1,bag2,bag3,bag4,bag5,bag6,bag7} need to make assumptions about the kernel, such as uniform blur, non-uniform blur, or depth unchanged, and then use the priors of natural images to restore a sharp image. Most of these traditional methods \cite{bag8,bag9,bag10,bag11,bag12} mainly focus on solving the motion blur caused by the motion of a simple object, camera translation, or other simple cases, while real blurred images are much more complicated. Therefore, it is still difficult for the traditional methods to solve the complex motion blur in real scenarios.

With great success in image classification and recognition, deep learning has also achieved breakthroughs in low-level tasks, such as single image super-resolution, JPEG artifacts removal, and denoising. Researchers have also begun to explore deblurring based on Convolutional Neural Networks (CNNs). Some early methods \cite{bag13,bag14,bag15,bag16} employ CNN to estimate the unknown blur kernels. Recently, some data-driven kernel-free end-to-end mapping methods \cite{bag17,bag18,bag19,bag20,bag21,bag22,bag23} were proposed and achieved significant progress. Inspired by the recent work on image super-resolution \cite{bag22} and GAN-based image style transfer \cite{bag23}, Kupyn \emph{et al}. \cite{bag20,bag21} addressed deblurring as a special case of image style transfer. A mapping network, from blurred images to sharp images, is trained, and then GAN is employed to recover the realistic, detailed textures, so as to improve visual quality. The end-to-end mapping methods mainly depend on paired training data. The deblurring network employs the same weights for different spatial positions of degraded images, so the method lacks the adaptive blurring-feature sensing mechanism to handle non-uniform blurred images.

The visual attention mechanism can locate the target area in an image and capture the features in the regions of interest. It has been successfully applied to the image classification and recognition problems \cite{bag24,bag25,bag26}. The goal of image deblurring is to restore the blurred regions of an image, therefore, we should pay more attention to the blurred regions. The main idea of our proposed method is to capture the spatially varying blur features in complex blurred images, using the spatial attention mechanism to guide the restoration.

The conditional GAN \cite{bag27} is employed as the backbone network, in which the generator realizes the mapping from complex blurred images to sharp images, and the discriminator distinguishes whether the image is generated or a ground truth. The blur-attention module is introduced into the generator to capture the spatially varying blur features. Fig. 1 shows the feature map of a non-uniform blurred image extracted using the blur-attention module. We can see that the feature map effectively captures the spatially varying blurring features, where the blurred regions are highlighted accurately. The main contributions of this paper are as follows.

\begin{itemize}
\item A blur-attention module is proposed. This module is comprised of a DenseBlock feature extraction unit and a multi-pooling feature-fusion-based spatial attention unit. A multi-level residual connection structure is designed and used to connect multiple modules into a blur-attention network, which can effectively capture the spatially varying blurring features in non-uniform blurred images.
\end{itemize}
\begin{itemize}
\item A blur-attention-based deblurring method for non-uniform blurred images is proposed. To the best of our knowledge, we are the first to introduce spatial attention mechanism for deblurring. This makes our proposed network have the ability to automatically capture the spatially varying blurring features.
\item By visualizing the feature maps generated by the attention module, the effectiveness of the attention module can be verified, which provides cues for the interpretability of the attention module. Comparative analysis with ResBlock-Attention and DenseBlock-Attention verify the superior performance of the proposed blur-attention feature perception module.
\item Comprehensive experiments are provided. The results show that our proposed blur-attention network can effectively boost the local adaptability of the algorithm. Both the subjective and objective quality of the restored images are improved significantly.
\end{itemize}

\begin{figure*}[!t]
 \begin{subfigure}{\linewidth} 
         \centering
        \includegraphics[width=  7cm]{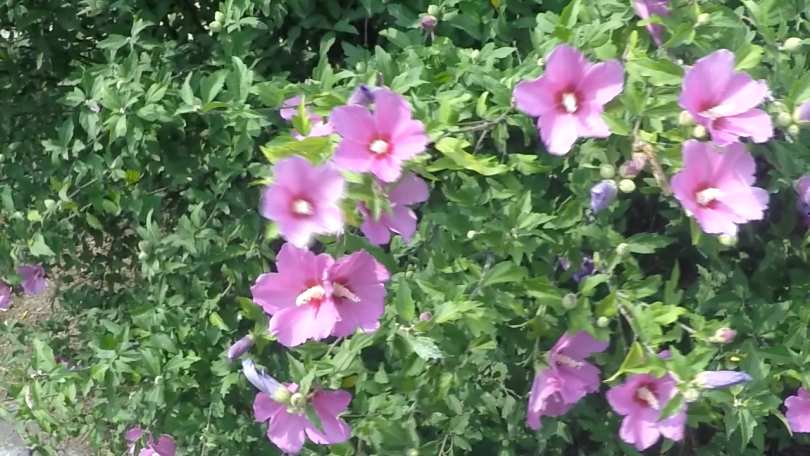}
        \caption{Blur image}
        \label{fig1a}
    \end{subfigure}
\\
  \begin{subfigure}{\linewidth}
        \centering
       \includegraphics[width= 7cm]{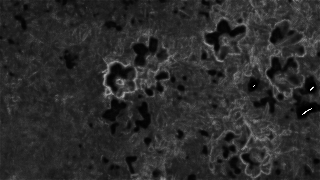}
       \caption{Blur feature map}
      \label{fig1b}
  \end{subfigure}  
\caption{A non-uniformly blurred image and its spatially varying blur feature map extracted by the blur-attention module.}
\label{fig1}
\end{figure*}

\section{Related work}

\subsection{Traditional image deblurring methods}
Deblurring has a long research history in image processing. The degradation model is defined as follows:

\begin{equation}
I_B = K \ast I_S + N,
\end{equation}
where $I_B$, $I_S$ and $N$ represent a vectorized blurred image, the sharp latent image, and noise, respectively. $''\ast''$ denotes the convolution operation, and $K$ is a matrix composed of local blur kernels, with each local blur kernel acting on the sharp image $I_S$ to generate a blurred pixel in $I_B$. In a real blurred image, the blur kernel is unknown, so this is blind restoration, which needs to estimate the blur kernel and the sharp latent image simultaneously. This is an ill-posed problem. Therefore, priors are applied to constrain the solution space in generating the sharp images from blurred images.

Early methods \cite{bag28,bag29} made the assumption that the blur kernel is spatially invariant. However, most of the real blurred images are subject to spatially varying kernels \cite{bag30}. Perrone \emph{et al}. \cite{bag31} assumed that the blur kernel is locally linear. The methods \cite{bag7} and \cite{bag28} employed an iterative strategy and a parameter prior model to improve the accuracy of blur kernel estimation for recovering the sharp images in successive iterations. However, the iterative procedure is usually time consuming, and these traditional methods have limitations in solving the restoration of blurred images.

\subsection{Learning-based image deblurring methods}

With the success of deep learning, CNN-based single image deblurring has been proposed recently. Early learning-based methods are generally divided into two steps. First, a deep neural network is used to estimate the blur kernel, and then a traditional deconvolution method is used to obtain a sharp latent image. Schuler \emph{et al}. \cite{bag15} used a coarse-to-fine approach to iteratively optimize blurred images, using multiple CNNs. Sun \emph{et al}. \cite{bag13} and Yan \emph{et al}. \cite{bag16} parameterized and estimated the blur kernel through classification and regression analysis. These methods \cite{bag13,bag15,bag16} follow the traditional framework, and use CNNs, instead of traditional methods, to estimate the unknown blur kernel. The deblurring performance of this type of algorithm depends on the accuracy of blur kernel estimation. Recently, end-to-end image deblurring, which does not require estimation of the blur kernel, has been proposed. This approach can reduce the errors caused by kernel estimation. Nah \emph{et al}. \cite{bag18} designed a multi-scale network to extract the multi-scale information of an image iteratively, which gradually restores the sharp image. Tao \emph{et al}. \cite{bag32} proposed a scaled recursive network model with shared parameters. However, these methods use the same network weights to recover blurred images, which lack an adaptive mechanism for handling non-uniform blurred images.

Most of the CNN-based image deblurring methods are end-to-end, data-driven methods, and the optimal cost function is usually based on objective indicators, such as average pixel difference (L1/L2 loss). Therefore, this type of method will improve the objective quality index of restored images, but their subjective quality may not be satisfactory, because of over-smoothing.

\subsection{Generative adversarial network}

Goodfellow \emph{et al}. \cite{bag1} proposed the Generative Adversarial Network (GAN), which consists of a generator and a discriminator. These two networks are trained in an adversarial manner. GAN has shown good results in some areas, such as image generation \cite{bag33} and completion \cite{bag34}. Pix2Pix \cite{bag23} employed GAN to achieve image style transfer and proposed a feature mapping method between input images and output images. Kupyn \emph{et al}. 
\cite{bag20,bag21} used GAN for the image deblurring task, using paired data for training, and achieved a good result. This shows the effectiveness of GAN in the image deblurring task. The adversarial network has changed the way of network training, and introduced the adversarial loss. The development of a series of new cost functions, such as perceptual loss and the images generated based on GAN have significantly improved the subjective quality of the restored images.

\subsection{Attention mechanism in the image restoration task}

The visual attention mechanism helps to locate the targets in an image and capture the features of the regions of interest. It has been successfully applied in the recognition and classification problems. At present, the attention mechanism is rarely used in image deblurring. The methods \cite{bag35} and \cite{bag36} applied the attention mechanism to the image restoration tasks for rain removal and multi-degradation restoration, respectively. The method in \cite{bag36} applied channel attention for the restoration of various types of degradation in an image. Different types of degradation, such as blur, compression distortion, and noise, select different filters to improve the robustness of the algorithm. As far as we know, currently there is no spatial attention-based deblurring task for non-uniform blurred images. The goal of non-uniform blurred image restoration is to restore the blurred regions of an image into a clear image. Therefore, the blurred regions in an image are the main area of attention for restoration, and the use of attention mechanism can effectively improve the performance for non-uniform image deblurring.

In this paper, we devise a blur-attention module, which can effectively locate the blurred regions in an image. This module first extracts features through the DenseBlock unit; followed by using the multi-pooling fusion spatial attention unit to dynamically capture the spatially varying features in non-uniform blurred images. Furthermore, we propose a multi-level residual connection structure to connect the multiple blur-attention modules to form a blur-attention network. Finally, the blur-attention network is introduced into a conditional GAN \cite{bag27} to dynamically restore the spatially non-uniform blurred images.

\section{Proposed method}
A novel blur-attention module is proposed to enable the deblurring network to adaptively capture the spatially varying blur features. We introduce the blur-attention module into the generator of a conditional GAN \cite{bag27}, to form the Blur-Attention-GAN (BAG) and an end-to-end method for dynamic scene deblurring. The framework of our proposed method is illustrated in Fig. 2. The generator adopts an encoder-decoder structure, and a blur-attention module is introduced between the encoding network and the decoding network. The generator restores the input blurred image to a sharp image. The discriminator distinguishes the generated image from the real high-quality sharp images. The generator and the discriminator are trained through a zero-sum game to achieve Nash equilibrium. In this section, first, we will address the blur-attention network. Then, we describe the details of the generator and discriminator. Finally, we will introduce the cost function.

\begin{figure*}[!t]
\centering
\includegraphics[width=\linewidth]{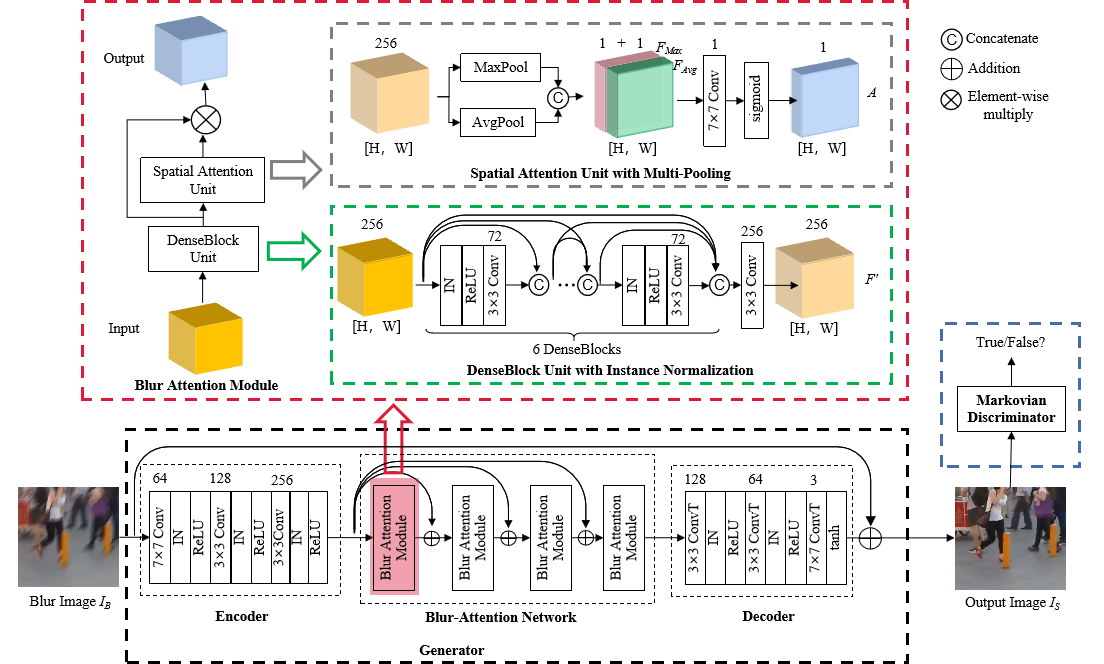}
\caption{The framework of Blur-Attention-GAN (BAG). The generator takes the blurred image as input and generates the restored sharp image. The discriminator is used to distinguish whether the restored image is real or fake. The blur-attention module adaptively captures the spatially varying blur features.}
\label{fig2}
\end{figure*}

\subsection{Blur-attention network}

The visual attention mechanism can locate the salient regions in an image that are the most discriminative for specific tasks. The idea of attention has been successfully applied to image recognition and classification. Fig. 1 is a typical blurred image in a dynamic scene. Most of the regions in the picture are sharp, but the shaking flowers are blurring, due to the blowing wind. The goal of a deblurring method is to restore the blurred parts of the image. Therefore, the deblurring network should deal with the blurred-flower regions and the sharp regions differently. The attention mechanism can effectively capture the spatially varying blur features, and consequently boost the visual quality of restored non-uniform blurred images.

We propose and implement a novel blur-attention network, which is composed of four new blur-attention modules, to extract spatially varying blur features. We also devise a multi-level residual connection structure to connect the four blur-attention modules. The blur-attention module and the multi-level residual connection are described in the following.

\textbf{Blur-attention module}. As shown in the upper left of Fig. 2, each blur-attention module has two units: DenseBlock Unit (DBU) and Spatial Attention Unit (SAU). Firstly, a DBU with Instance Normalization (IN) is employed for feature extraction. Six DenseBlocks are cascaded to form a DBU. The SAU is employed to generate a 2D spatially varying blur-feature map by the multi-pooling fusion. The blur-feature map and the network output feature map are multiplied element-wise to realize the spatial attention feature selection mechanism. The process of the blur-attention module is shown as follows:

\begin{equation}
    \begin{aligned}
        F_{out} &= Net_{BAM}(F_{in})\\
          &= Net_{SAU}(Net_{DBU}(F_{in}))\otimes Net_{DBU}(F_{in}),\\
          &=A\otimes F'
    \end{aligned}
\end{equation}
where $\otimes$ represents element-wise multiplication, $F_{in}$ and $F_{out}$ denote the input and output, respectively, of the blur-attention module, and $Net_{BAM}$, $Net_{SAU}$, and $Net_{DBU}$ denote the output of the blur-attention module, the spatial attention unit, and the DenseBlock unit. $A$ and $F'$ represent the 2D attention feature map extracted by the SAU and the output feature map by the DBU, respectively. The following is a detailed introduction to DBU and SAU.

(1) \textbf{DenseBlock Unit (DBU)}. To extract multi-layer fusion features from the input feature map, Instance Normalization is employed to normalize the feature map of each training image. Unlike DeblurGAN \cite{bag20}, which uses a series of residual blocks to extract features, we employ an enhanced DenseBlock to extract features, which is an improved version of the original DenseBlock \cite{bag37}. The original DenseBlock is designed for classification, which is different from low-level tasks. We replace batch normalization (BN) in the original DenseBlock to Instance Normalization (IN). Image deblurring can be seen as a mapping between a blurred image and the corresponding clear image. Each blurred image has its own non-uniform blur characteristics, and the specific blur kernel function for different spatial locations is different. Therefore, instance normalization is selected as the normalization operation of the entire network, in order to maintain the independence between two different image instances. The normalization algorithm normalizes the input image according to its specific characteristics, regardless of other training samples.

In the experiments, each DBU is composed of 6 convolutional layers. The kernel size of each convolutional layer is 3$\times$3. The number of output channels is 72, and all the activation functions used are ReLU. The output feature maps, via different convolutional layers, are concatenated, as shown in Fig. 2.

(2) \textbf{Spatial Attention Unit (SAU)}. To capture the spatially varying blur features of a non-uniform blurred image, we introduce the spatial attention mechanism. Unlike a binary mask, our attention map is a matrix with element values between 0 and 1. The larger the value, the higher the importance is, and the more attention should be paid to its corresponding element.

The SAU employs a multi-pooling feature fusion strategy to help the network to extract the spatially varying blur features. The output $F' \in  R^{C\times H\times W}$ of the DBU is used as the input of the SAU, and a 2D attention mask $A \in  R^{1\times H\times W}$ is produced, as shown in Fig. 2. First of all, max pooling and average pooling are applied to the input feature maps in the channel dimension. The features are extracted from every spatial position of each feature map, and the two features, $F_{Max} \in  R^{1\times H\times W}$ and $F_{Avg} \in  R^{1\times H\times W}$, are extracted. Then, the pooling results are concatenated together. Finally, a convolution operation is used to obtain the 2D attention mask $A \in R^{1\times H\times W}$. Among the multi-pooling operation, max pooling and average pooling can retain the edge and background information, respectively, in the features. The splicing of these two kinds of features may provide rich information, and ensure that the attention network can accurately extract the spatial information of the input image. The specific process is as follows.

\begin{equation}
    \begin{aligned}
        A &= \sigma(f[MaxPool(F'),AvgPool(F')])\\
          &= \sigma(f[F'_{Max},F'_{Avg}]),
    \end{aligned}
\end{equation}
where $ A$ is the attention feature map extracted by the SAU. $F'$ is the output feature map of DBU. $ f$ represents the convolution operation with a kernel of size 7$\times$7, and $ \sigma$ is the sigmoid activation function. [,] represents the concatenation operation. With pooling channel-wise, the number of channels of the feature maps is reduced, and the spatial resolution remains unchanged. Therefore, multiple blur-attention modules are connected, and the spatial resolution of the feature maps remains unchanged.

\textbf{Multilevel residual connection}. The residual network \cite{bag40} can help a deep network to converge faster. Both DeepDeblur \cite{bag18} and DeblurGAN \cite{bag20} employ the residual blocks to build their feature transformation module. Specifically, the operation of a residual block is defined as follows:

\begin{equation}
y = F(x)+x,
\end{equation}
where $x$ and $y$ denote the input and output of the residual module, and $F$ represents the residual function. We name this connection method as a one-level residual connection, as shown in Fig. 3 (a).

\begin{figure*}[!t]
\centering
 \begin{subfigure}{0.4\linewidth}
        \includegraphics[width= 3.0cm]{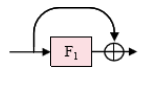}
        \caption{One-level residual connection}
    \end{subfigure}
    \begin{subfigure}{.4\linewidth}
        \includegraphics[width= 4.45cm]{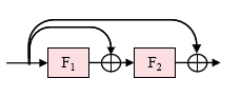}
        \caption{Two-level residual connection}
    \end{subfigure}
    \begin{subfigure}{\linewidth}
    \centering
        \includegraphics[width= 6.65cm]{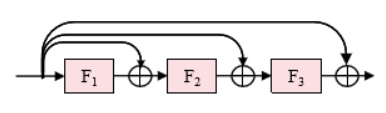}
        \caption{Three-level residual connection}
    \end{subfigure}
\caption{Multilevel residual connections.}
\label{fig3}
\end{figure*}

Figs. 3(b) and 3(c) show the two-level residual connection and the three-level residual connection, respectively. Unlike the ordinary one-level residual connection, multi-level residual connections add the input of the first residual module and the output of the previous residual module to form the input of the next module. Multilevel residual connections can reduce the loss of low-level information in the transmission process and accelerate the convergence of the network. Although multiple cascaded residual modules contain multiple short skip connections, some information will still be lost in the process of unidirectional transmission. The input of each level of our multilevel residual connection contains the input information of the first level, so that each module can extract the difference information between the input and the output. The residual information extracted by the previous module can assist the next module in better learning the varying characteristics from blur to sharp features. We believe that these differential features are blur dynamic features, and our multi-level residual connection helps the blur-attention module to better extract the spatially varying blur features gradually.

\subsection{Conditional GAN for image deblurring}

We introduce the blur-attention module into a conditional GAN \cite{bag27}, and propose the Blur-Attention-GAN (BAG) method. As shown in Fig. 2, the generator is based on an encoder-decoder structure, whose input is a blurred image and the output is a sharp latent image. Our generator is composed of three parts: an encoder network, a blur-attention network, and a decoding network. The encoding network consists of three convolutional layers, and the decoding network consists of three transposed convolutions. All the convolutional layers of the encoder and decoder adopt IN and ReLU.

We include a global long connection to add the network input to the output to facilitate network convergence. Unlike the method in \cite{bag20}, which employs a series of stacked residual blocks to form a non-linear transformation network, we design a multi-level residual connection structure to connect the multiple blur-attention modules to form a blur-attention network. This network can automatically extract the spatially varying blur features.

\subsection{Discriminator}

A non-uniform blurred image contains both sharp regions and blur regions. The traditional GAN employs the global content consistency loss and detects whether an entire image is real through a global discriminator, which does not consider the local characteristics. Our discriminator employs the Markov discriminant network \cite{bag38}, which is used to enhance the extraction and characterization of local features.

The network structure of our discriminator is the same as PatchGAN \cite{bag23}. The discriminator is composed of a fully convolutional network. Each element in the output feature map represents a patch in the input image. This discriminator can help the network to better restore the image details.

\subsection{Loss functions}

Given a pair, consisting of a blurred image and a sharp image, the loss function used in the training process of the network is similar to DeblurGAN \cite{bag20}. The content loss function and the adversarial function form the joint loss of the network, as follows:

\begin{equation}
L = L_{GAN}+\lambda L_{C},
\end{equation}
where $L_{GAN}$ and $L_C$ represent the adversarial loss and the content loss, respectively. Same as the setting for DeblurGAN, $\lambda$ is set at 100 in the experiments.

The Wasserstein-GAN \cite{bag41} is selected as the adversarial loss function. The original loss function of the GAN is gradient instability, which makes the network difficult to train and converge. The Wasserstein-GAN is more robust. The adversarial function is shown as follows:

\begin{equation}
L_{GAN}=\min_G\max_D \sum_{i=1}^N(D_{\theta_D}(I_S)-D_{\theta_D}(G_{\theta_G}(I_B))),
\end{equation}
where $I_S$ and $I_B$ represent the sharp image and the blurred image, respectively, and $N$ is the number of training images.

We select the perceptual loss \cite{bag42} as the content loss. The perceptual loss is a simple L2 loss, which is calculated based on the difference between the features generated by VGG19 from the generated image and the target image. The loss function based on the feature from the ith convolutional layer of the pre-trained VGG19 is given as follows:

\begin{equation}
L_C=\frac{1}{C_iW_iH_i} \sum_{z=1}^{C_i}\sum_{x=1}^{W_i}\sum_{y=1}^{H_i}(\phi_i(I_S)_{x,y,z}-\phi_i(G_{\theta_G}(I_B))_{x,y,z})^2,
\end{equation}
where $C_i$, $W_i$, and $H_i$ represent the dimensions of the feature map at the $i$th convolutional layer in the pre-trained VGG19 network. In our experiments, we choose $i$ = 7.

\section{Experimental results}

We conducted intensive experiments to evaluate our proposed blur-attention module. We will first introduce the dataset used and the experiment settings, ablation experiments for the blur-attention module, and comparisons with other similar methods. To make a fair comparison, all the compared methods are trained and tested on the same dataset. The source codes of the comparison methods are released by the corresponding authors, with the default parameters. The platform used for our experiments is NVIDIA GeForce GTX 1080 Ti GPU.

\subsection{Data and experimental settings}

Previous learning-based methods usually applied blur kernels to sharp images to simulate the blurred images. It is difficult to generate complex non-uniform blurred images as real blurred images. In \cite{bag18}, the consecutive frames in a video captured by a high-speed camera are averaged to synthesize real blurred images, and the GoPro dataset was created. The dataset contains a variety of scenes, which can well simulate complex camera shake and object motion.

\textbf{Dataset}. In order to make a fair comparison with other methods, all the methods compared in the experiments are trained based on the GoPro dataset \cite{bag18}. This dataset contains 3,214 pairs of blurred and sharp images. Following the same strategy in \cite{bag18}, we selected 2,103 pairs of images as the training set, and the remaining 1,111 pairs of images as the testing set.

\textbf{Model parameter settings}. When training our network model, the batch size is set at 1, and the Adam optimization algorithm is selected, where  $\beta_1$ = 0.9 and $\beta_2$ = 0.999. The initial learning rate of the generator and the discriminator are both set at $10^{-4}$, and the epoch is set to 300. In the last 150 epochs, the learning rate is linearly degraded to zero. To balance the generator and the discriminator, the discriminator is updated 5 times during training, then the generator is updated once. We randomly crop patches of size 256$\times$256 pixels from the original images to form the training input.

\subsection{Ablation experiments}

The blur-attention network is the major contribution of this paper. We conducted extensive ablation experiments to evaluate the performance of the blur-attention network. The average PSNRs for different configurations of the blur-attention network are measured, and listed in Table 1. The analysis of the results provides inspiration for the design of the blur-attention network.

\subsubsection{Effectiveness of the blur-attention module}

The blur-attention module is composed of DBU and SAU. To evaluate the effectiveness of DBU with IN, we compare several basic structures, with DBU replaced. Six different structures are considered and evaluated. As shown in Table 1, "Model Plain" is the structure with DBU replaced by 7 convolutional layers. "Model 1" uses ResBlock model, instead of DBU, "Model 2" employs DenseBlock with Batch Normalization, and "Model 3" uses DenseBlock with Instance Normalization, as our proposed DBU. These 4 models are used to analyze the effectiveness of DBU.

In order to make a fair comparison, the number of convolutional layers used as replacement is the same as the blur-attention module, and the other parts are consistent with the BAG. All the modules have 7 convolutional layers.

It can be seen from Table 1 that the PSNR performance of Model 1 and Model 2 is better than that of Model Plain, which employs convolutional layers only. This indicates that the feature-extraction ability of the residual block and the DenseBlock is more powerful than the plain convolutional layers. Model 3 achieves a higher PSNR than the previous three models. This implies that Instance Normalization improves our model by 0.6dB, compared to BN, in terms of the average PSNR. We can see that our proposed DBU (Model 4) brings about a 2dB gain compared to the plain convolutional feature-extraction model.

\begin{table}[!t]
\renewcommand{\arraystretch}{1.3}
\caption{The average PNSR performances for different structures of the blur-attention module in the ablation experiments.}
\label{table_I}
\centering
\setlength{\tabcolsep}{1.3mm}{
\begin{tabular}{c|c|c|c|c|c|c}
\hline\hline

\multirow{2}{*}{Model} &Model &Model &Modle&Modle &Modle & \multirow{2}{*}{BAG} \\
 &Plain& 1 & 2 & 3 &  4 &  \\
\hline
ResBlock                          &  &\checkmark & &  &  &  \\
DenseBlock-BN                     &            & &\checkmark &  &  &  \\
DenseBlock-IN                     &            & & &\checkmark  &\checkmark  &\checkmark \\
SAU                               &            & & &   &\checkmark  &\checkmark  \\
Residual connection               &            & & &  &\checkmark	 &  \\
Multilevel residual connection    &\checkmark  &\checkmark &\checkmark &\checkmark  & &\checkmark \\
PSNR(dB)                          &27.08       &28.20 &28.80 &29.06  &29.14	 &29.41 \\
\hline\hline
\end{tabular}}
\end{table}

To evaluate the performance of SAU, our BAG method is the Model 3 with SAU. It can be seen from Table 1 that the PSNR of the BAG model is further improved from 29.06 dB to 29.41dB. We can conclude that SAU improves the average PSNR by 0.35dB.

\subsubsection{The effectiveness of the multi-level residual connections}

In order to prove the effectiveness of the multi-level residual connection, Model 4 employs the residual connection in the blur-attention network. Compared with the residual connection, our multi-level residual connection model improves the PSNR from 29.14dB to 29.41 dB, i.e. a gain of 0.27dB.

\subsubsection{Subjective comparisons}

The subjective results of the ablation experiments are shown in Fig. 4. From (a) to (f), the blurred image, followed by the results of the Model Plain, the ResBlock (Model 1), the DenseBlock (Model 3), our BAG, and ground truth, are illustrated.

As shown in Fig. 4, both scene blur and motion blur exist in the blurred image, which may be caused by the joint action of the camera movement and the scene target. The Model Plain improves the overall clarity of the image. However, the restoration of the scenes with different depths has different degrees of clarity, such as the leg of the walking pedestrians. The edges of the static yellow stick are not sharp enough. Some artifacts appear near the top and the boundary of the sticks. From the image point of view, it seems that the same operation is applied to the non-uniform blurred regions. The result of ResBlock is better than Model Plain. The pedestrians with different distances are still blurred, and the second yellow stick still has a blurred boundary. For the result based on DenseBlock, the artifacts around the yellow sticks have been improved significantly. The purple backpack is also clear. However, there are still obvious differences near the lags of different pedestrians. This is due to the network's lack of adaptive perception of blur features at different depths. These methods tend to use the uniform manner to restore the global image. In our BAG method, the blur-attention mechanism perceives the changes in blur in the spatial position, and guides the network to recover the blurred images adaptively. We can see that the legs of the pedestrians at different depths can be significantly restored. The yellow static sticks are also very clear. It can be seen from the figure that the result from BAG is the most similar to the ground truth.

\begin{figure*}[!t]
\centering
 \begin{subfigure}{0.3\linewidth}
        \includegraphics[width= 3.5cm]{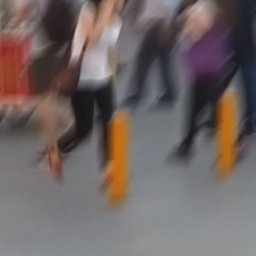}
        \caption{Blurred image}
    \end{subfigure}
 \begin{subfigure}{0.3\linewidth}
        \includegraphics[width= 3.5cm]{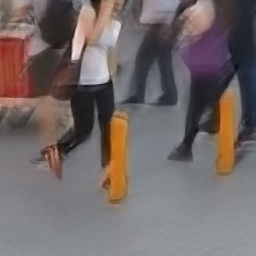}
        \caption{The result of Model Plain}
    \end{subfigure}
 \begin{subfigure}{0.3\linewidth}
    \centering
        \includegraphics[width= 3.5cm]{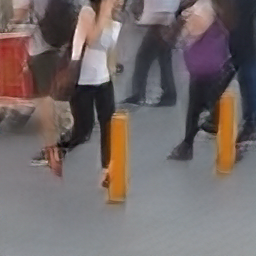}
        \caption{The result of ResBlock}
    \end{subfigure}
 \begin{subfigure}{0.3\linewidth}
    \centering
        \includegraphics[width= 3.5cm]{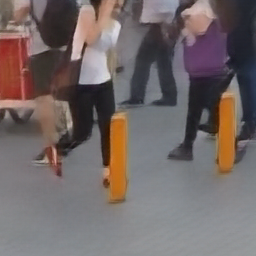}
        \caption{The result of DenseBlock}
    \end{subfigure}
 \begin{subfigure}{0.3\linewidth} 
    \centering
        \includegraphics[width= 3.5cm]{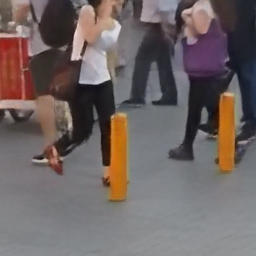}
        \caption{The result of BAG}
    \end{subfigure}
 \begin{subfigure}{0.3\linewidth}
    \centering
        \includegraphics[width= 3.5cm]{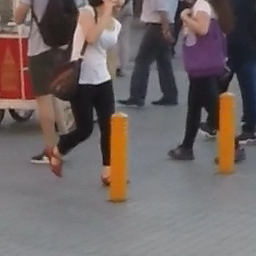}
        \caption{Ground truth}
    \end{subfigure}
\caption{Subjective results of the ablation experiments.}
\label{fig4}
\end{figure*}

\subsection{Comparisons with other deblurring methods}

In order to evaluate the effectiveness of our method, we compare it with several state-of-art image deblurring methods on different datasets, including a real blurred dataset. These methods include Sun's method \cite{bag39}, Nah's method \cite{bag18}, DeblurGAN \cite{bag20}, DeblurGAN-v2 \cite{bag21}, and our BAG method. Sun's method \cite{bag39} is the first method that uses CNN to estimate the blur kernel. The method of Nah \emph{et al}. \cite{bag18} employs an end-to-end network to restore the image and achieves good results. DeblurGAN \cite{bag20} applies GAN to image deblurring tasks, and this method can better restore image details. DeblurGAN-v2 \cite{bag21} is the latest proposed image deblurring algorithm, which has achieved state-of-the-art performance on the GoPro dataset.

\subsubsection{Results on the GoPro dataset}

Our method and the current non-uniform deblurring method were experimentally tested on the GoPro evaluation dataset. Table 2 tabulates the objective PSNR and SSIM experimental results of different methods. The experimental results show that Sun \cite{bag39} and Nah 's methods \cite{bag18} can achieve improvements on the GoPro dataset. Our method using DeblurGAN \cite{bag20} as the baseline, with the proposed blur-attention module added, improves the PSNR by 2.2 dB, which brings significant visual quality improvement. Our model can handle blurred images in highly dynamic scenes well. The PSNR of our method can achieve PSNR of 29.4dB, which is comparable to the results of DeblurGAN-v2.

DeblurGAN-v2 \cite{bag21} introduces a feature pyramid network to fully extract the priori information of blurred images at different scales. In order to improve the performance, a complex inception-ResNet-v2 backbone network is employed. These strategies increase the computational complexity. To deal with this issue, a MobileNet-DSC \cite{bag21} model was introduced, but it results in a decrease in PSNR by about 1.5dB.

The computational costs of the different methods are listed in Table 2. We can see that our method is much faster than DeblurGAN-v2 \cite{bag21}. The average runtime required for an image in the GoPro dataset, based on DeblurGAN-v2 \cite{bag21}, is 7.15s, while our method only requires 1.13s.

\begin{table}[!t]
\renewcommand{\arraystretch}{1.3}
\caption{The average PSNRs, SSIMs and runtimes on the GoPro dataset.}
\label{table_II}
\centering
\begin{tabular}{l|c|c|c}
\hline\hline
Method&PSNR(dB)&SSIM&Time \\
\hline
Sun \emph{et al}. \cite{bag39}&24.6 & 0.84 &20 min \\
\hline
DeepDeblur \cite{bag18}&29.1 & 0.91 & 3.09 s \\
\hline
DeblurGAN \cite{bag20}&27.2 & 0.95 & 0.97 s \\
\hline
DeblurGAN-v2 \cite{bag21}&29.6 & 0.93 & 7.15s \\
\hline
Our method &29.4 & 0.89 & 1.13 s \\
\hline\hline
\end{tabular}
\end{table}

The subjective performance of the different methods is shown in Fig. 5. From the two zooming-in regions of the image on the left column, one region is the faraway license plate, and the other one is a region on the ground. From the original image, we can see that the appearance of the license plate changes gradually from blurry to clear and legible. From the deblurred results based on the different methods, the digits on the license plate generated by our method are the clearest, and also the closest to the clear image of the target shown in the last row. The texture artifacts on the ground are also the least based on our method, compared to other methods. The restoration results show that our method can almost eliminate all the artifacts. Experimental results show that our method can handle blur regions with different depths at the same time. This is benefitted from the blur-change perception ability of the blur-attention module.

In the image on the right column, the two zoom-in regions are a static light around the top and a dynamic pedestrian’s hand, respectively. It can be seen from the experimental results that the strip texture in the static-light region, based on our method, is different from the restoration results by the other methods. The restoration results of our method are the most similar to the target image, shown in the bottom. In the motion-blur areas of the pedestrian's hand, our restored result is also the closest to the clear result of the target. This shows that our blur-attention module can handle both static and dynamic motion regions effectively and simultaneously.

\begin{figure*}[!t]
\centering
 \begin{subfigure}{0.45\linewidth}
        \includegraphics[height=2.5cm,width=5.42cm]{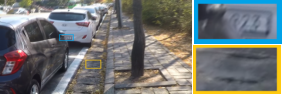}
    \end{subfigure}
 \begin{subfigure}{0.45\linewidth}
        \includegraphics[height=2.5cm,width=5.42cm]{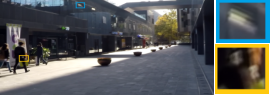}
    \end{subfigure}
 \begin{subfigure}{0.45\linewidth}
        \includegraphics[height=2.5cm,width=5.42cm]{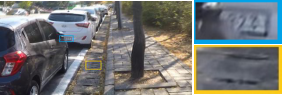}
    \end{subfigure}
 \begin{subfigure}{0.45\linewidth}
        \includegraphics[height=2.5cm,width=5.42cm]{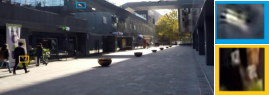}
    \end{subfigure}
 \begin{subfigure}{0.45\linewidth}
        \includegraphics[height=2.5cm,width=5.42cm]{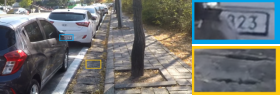}
    \end{subfigure}
 \begin{subfigure}{0.45\linewidth}
        \includegraphics[height=2.5cm,width=5.42cm]{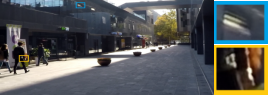}
    \end{subfigure}
\begin{subfigure}{0.45\linewidth}
        \includegraphics[height=2.5cm,width=5.42cm]{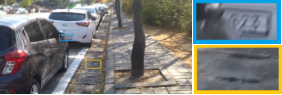}
    \end{subfigure}
 \begin{subfigure}{0.45\linewidth}
        \includegraphics[height=2.5cm,width=5.42cm]{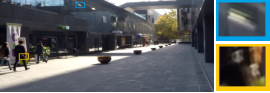}
    \end{subfigure}
 \begin{subfigure}{0.45\linewidth}
        \includegraphics[height=2.5cm,width=5.42cm]{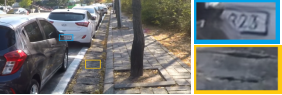}
    \end{subfigure}
 \begin{subfigure}{0.45\linewidth}
        \includegraphics[height=2.5cm,width=5.42cm]{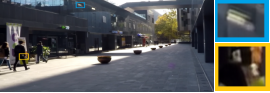}
    \end{subfigure}
 \begin{subfigure}{0.45\linewidth}
        \includegraphics[height=2.5cm,width=5.42cm]{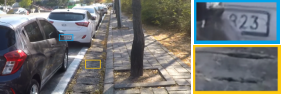}
    \end{subfigure}
 \begin{subfigure}{0.45\linewidth}
        \includegraphics[height=2.5cm,width=5.42cm]{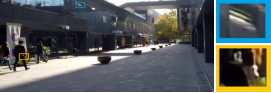}
    \end{subfigure}
\begin{subfigure}{0.45\linewidth}
        \includegraphics[height=2.5cm,width=5.42cm]{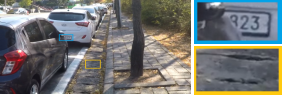}
    \end{subfigure}
 \begin{subfigure}{0.45\linewidth}
        \includegraphics[height=2.5cm,width=5.42cm]{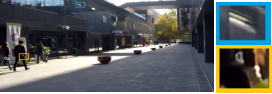}
    \end{subfigure}
\caption{Subjective results of different methods on two images from the GoPro dataset. From top to bottom, the blurred image, the results by Sun \emph{et al}.  \cite{bag39}, DeepDeblur \cite{bag18}, DeblurGAN \cite{bag20}, DeblurGAN-v2  \cite{bag21}, our method, and the ground-truth image are shown.}
\label{fig5}
\end{figure*}

\subsubsection{Results on real blurred images}

The GoPro test data is synthesized by a high-speed camera, which is still different from the real blurred image. To evaluate the robustness of our model, we test different methods on real blurred images, downloaded from the Internet, as shown in Fig. 6. Our trained model generalizes well on these images. Compared with DeblurGAN \cite{bag20} and DeblurGAN-v2 \cite{bag21}, our results are of higher visual quality, and our method restores visual details better than other methods.

\begin{figure*}[!t]
\centering
 \begin{subfigure}{0.45\linewidth}
        \includegraphics[height=2.5cm,width=5.42cm]{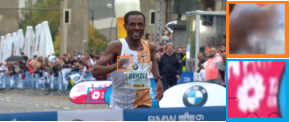}
    \end{subfigure}
 \begin{subfigure}{0.45\linewidth}
        \includegraphics[height=2.5cm,width=5.42cm]{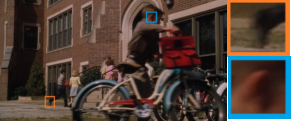}
    \end{subfigure}
 \begin{subfigure}{0.45\linewidth}
        \includegraphics[height=2.5cm,width=5.42cm]{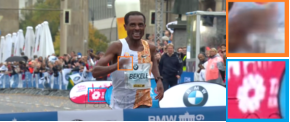}
    \end{subfigure}
 \begin{subfigure}{0.45\linewidth}
        \includegraphics[height=2.5cm,width=5.42cm]{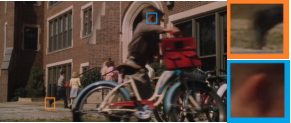}
    \end{subfigure}
 \begin{subfigure}{0.45\linewidth}
        \includegraphics[height=2.5cm,width=5.42cm]{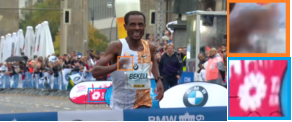}
    \end{subfigure}
 \begin{subfigure}{0.45\linewidth}
        \includegraphics[height=2.5cm,width=5.42cm]{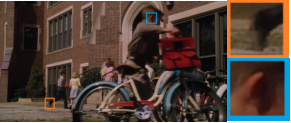}
    \end{subfigure}
\begin{subfigure}{0.45\linewidth}
        \includegraphics[height=2.5cm,width=5.42cm]{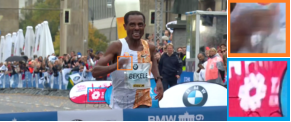}
    \end{subfigure}
 \begin{subfigure}{0.45\linewidth}
        \includegraphics[height=2.5cm,width=5.42cm]{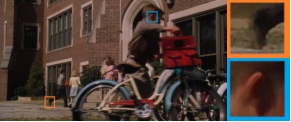}
    \end{subfigure}
\caption{Real blurred images. From top to bottom are the blurred image, DeblurGAN \cite{bag20}, DeblurGAN-v2 \cite{bag21} test result, and the result of our method.}
\label{fig6}
\end{figure*}

\section{Discussions}

An extensive discussion of our proposed blur-attention model is made in this section in order to provide further insights into the proposed blur-attention model and its potential future development.

\subsection{The role of attention modules in different positions}

We employ four blur-attention modules in the blur-attention network. In order to explore the role of attention modules in different positions, we visualized the output feature maps of the different attention modules. Figure 7 shows the visualization results of the attention maps extracted by the four blur-attention modules on a test image. It can be seen from the results that the attention weights from the four blur-attention modules, from the front-to-back blur areas (i.e., from $A^1$ to $A^4$), gradually increase, while the attention weights for the clear background gradually decrease. It is worth noting that the attention weights of the blurred area are not the same as the blur-attention modules, because the blurriness of the non-uniform blurred image spatially varies, and the attention weights for the regions around blurred scenes are relatively higher.

\begin{figure*}[!t]
\centering
 \begin{subfigure}{0.4\linewidth}
        \includegraphics[height=3.36cm,width= 4.8cm]{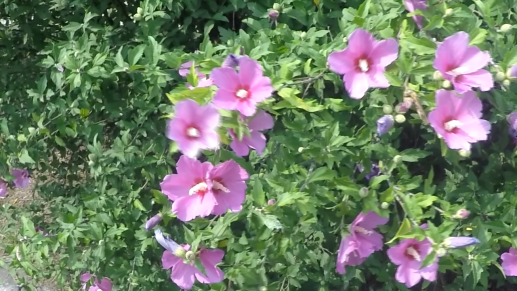}
        \caption{Blurred image}
    \end{subfigure}
 \begin{subfigure}{0.4\linewidth}
        \includegraphics[height=3.36cm,width= 4.8cm]{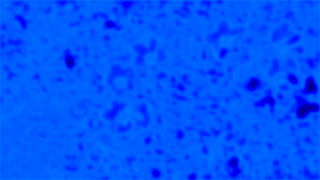}
        \caption{$A^1$}
    \end{subfigure}
 \begin{subfigure}{0.4\linewidth}
    \centering
        \includegraphics[height = 3.36cm,width= 4.8cm]{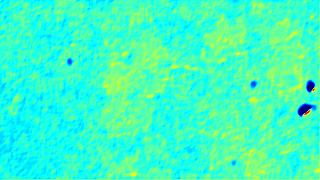}
        \caption{$A^2$}
    \end{subfigure}
 \begin{subfigure}{0.4\linewidth}
    \centering
        \includegraphics[height = 3.36cm,width= 4.8cm]{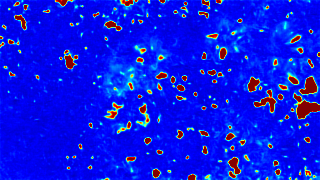}
        \caption{$A^3$}
    \end{subfigure}
 \begin{subfigure}{1\linewidth} 
    \centering
        \includegraphics[height = 3.36cm,width= 4.8cm]{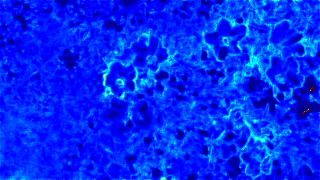}
        \caption{$A^4$}
    \end{subfigure}

\caption{Visualized outputs of the attention feature maps $A^i$ on a test image. Cool colors represent lower weights, and warm colors indicate higher weights.}
\label{fig7}
\end{figure*}

\subsection{Visualization of an attention map during training}

We also visualized the feature map of the fourth attention module, i.e., $A^4$, during training after different numbers of epochs, as shown in Fig. 8. The visualization results show that the network gradually pays more attention to the blurred areas during the training process, and the focus on the clear background gradually reduces, indicating that the proposed blur-attention model helps the network to extract the spatially varying blur features effectively.

\begin{figure*}[!t]
\centering
 \begin{subfigure}{0.3\linewidth}
        \includegraphics[height = 2.5cm,width= 3.5cm]{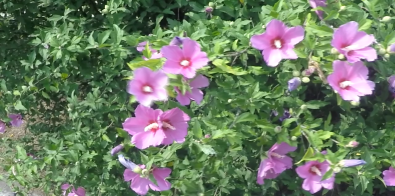}
        \caption{Input}
    \end{subfigure}
 \begin{subfigure}{0.3\linewidth}
        \includegraphics[height = 2.5cm,width= 3.5cm]{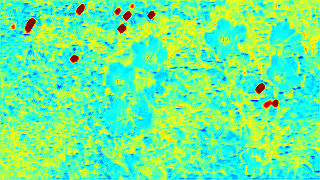}
        \caption{epoch=5}
    \end{subfigure}
 \begin{subfigure}{0.3\linewidth}
    \centering
        \includegraphics[height = 2.5cm,width= 3.5cm]{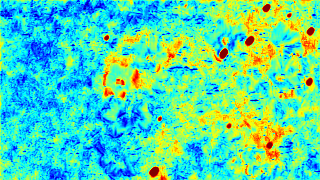}
        \caption{epoch=50}
    \end{subfigure}
 \begin{subfigure}{0.3\linewidth}
    \centering
        \includegraphics[height = 2.5cm,width= 3.5cm]{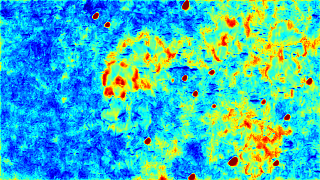}
        \caption{epoch=100}
    \end{subfigure}
 \begin{subfigure}{0.3\linewidth} 
    \centering
        \includegraphics[height = 2.5cm,width= 3.5cm]{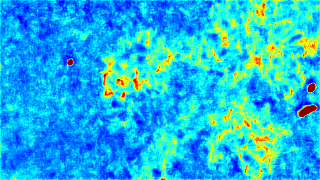}
        \caption{epoch=150}
    \end{subfigure}
 \begin{subfigure}{0.3\linewidth}
    \centering
        \includegraphics[height = 2.5cm,width= 3.5cm]{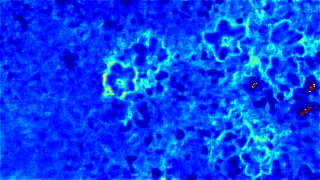}
        \caption{epoch=200}
    \end{subfigure}
\caption{The visualization of the fourth attention feature map $A^4$ during the training process. In the attention map, warm colors represent larger values, and cool colors represent smaller values. It can be seen that the weights for the clear background gradually decrease during training, and the weights for the blurred areas gradually increase.}
\label{fig8}
\end{figure*}

\subsection{Other discussions}

The blur-attention mechanism enables the network to dynamically capture spatially varying blur features, so as to guide the restoration network to produce clear and high-quality restored images. From the subjective experiments shown in Fig. 5, we can see that the blur-attention mechanism can handle the different blurring issues, arising from different depths, motions, and static defocus. The adaptive power provided by the blur-attention mechanism has great potential to deal with the complex deblurring problem.

However, this method can still be further improved for the restoration of real blurred images. In training our current model, we employed a limited number of real training samples, and focused on considering deblurring only. The degradation process in real scenarios should be more complicated. Therefore, multi-degradation factors may be coupled, so that even higher-quality images can be restored.

The attention mechanism is of particular importance in solving the problem of image deblurring in dynamic scenes. We believe that the selection of the backbone network and the location of the attention modules also play an important role in the deblurring performance. In DeblurGAN-v2 \cite{bag21}, a more complicated pyramid backbone brings significant improvement in the objective index, which is comparable with our boosting results on the plaint encoder-decoder backbone. All these will be explored in our future research.

\section{Conclusion}

In this paper, we propose an end-to-end deblurring method for dynamic scenes. We have proposed a blur-attention module to the conditional GAN, so that the network can adaptively capture the spatially varying blur features, and can dynamically restore blurred images. This method first applies GAN with a visual attention mechanism to the image deblurring task. By visualizing the attention map, the effectiveness of the attention module can be verified. Experimental results prove that this method has achieved promising subjective and objective performance.

\bibliography{BlurAttention}

\end{document}